\documentclass{kapproc} 
\usepackage{graphics}

\kluwerbib

\begin{document}

\articletitle{Three aspects of red giant studies in the Magellanic Clouds}

\articlesubtitle{}

\author{Maria-Rosa L. Cioni}
\affil{European Southern Observatory\\
Karl--Schwarzschild--str. 2,  85748 Garching bei Muenchen (Germany)}
\email{mcioni@eso.org}

\chaptitlerunninghead{Red giants in the Magellanic Clouds}

\begin{abstract}
There  are three  important aspects  concerning the  study of  the red
giant and in particular of  the asymptotic giant branch (AGB) stars in
the Magellanic Clouds. These are: the surface distribution, the luminosity 
function and the variability. The  spatial
distribution of AGB stars is  an efficient tool to study the structure
of the galaxies  and their metalicity by analysing  the ratio between
carbon--  and  oxygen--rich AGB  stars. The shape of the luminosity function
carries informations about  the star formation rate in  the Clouds and
it  can  be mathematically  related  to  their  history.
   Most  AGB  stars vary  their
magnitude in a few to several  hundred years time; the one epoch DENIS
magnitudes for both Large and Small Magellanic Cloud AGB stars outline
the same relations as a function of period.
\end{abstract}

\begin{keywords}
Magellanic Clouds, red variables, massive photometry.
\end{keywords}

\section{Introduction}
Studying  the  stellar  content  of  the Magellanic  Clouds  has  some
advantages: these  are nearby galaxies  that can be fully resolved in stars,  
they are relatively un--obscured and all their stars are
at about the  same distance. Before the publication  of the catalogues
and follow--up studies  from large scale surveys what  was known about
red  giants  in  the  Clouds  were  spatially  and  magnitude  limited
informations,  often  focalised   on  particular  objects  (i.e.  Mira
variables).   The past  few  years  have seen  the  release of the  DENIS
catalogue   towards  the   Magellanic   Clouds  (DCMC   --  Cioni   et
al.~\cite{cioni1}) which provides photometry in the broad $I$, $J$ and
$K_S$ bands.  These filters are particularly suitable for the study of
the red  giant population (Cioni  et al.~\cite{cioni2}, \cite{cioni3},
\cite{vdm1}).   The  release  of  the  2MASS  catalogue  (Nikolaev  \&
Weinberg \cite{2mass}) provides photometry in the  $J$, $H$ and
$K_S$  broad bands and  reaches slightly  fainter magnitudes  than the
DENIS measurements. The three micro-lensing surveys OGLE, MACHO \& EROS
provide light--curves over several years in two very broad blue
and red filters that allow  to characterise the complex variability of
the red giants,  as discussed also in the contribution by Wood (this 
proceeding).  The
OGLE  and MACHO  datasets are  at present  publically  available.  The
combination of the near--infrared  observations and light--curves is a
key  to understand  the  evolution and  properties of  long--period
variables.   In    addition   Zaritsky   (\cite{zari}) published his $UBVI$ 
survey of the SMC providing also an extinction map of the Cloud. Finally, 
 Massey (\cite{massey}) presented less sensitive $UBVR$ measurements in 
both Clouds to characterise the more massive and brighter stellar 
populations.  
\begin{figure}
\begin{minipage}[t]{6cm}
\resizebox{6cm}{!}{\includegraphics{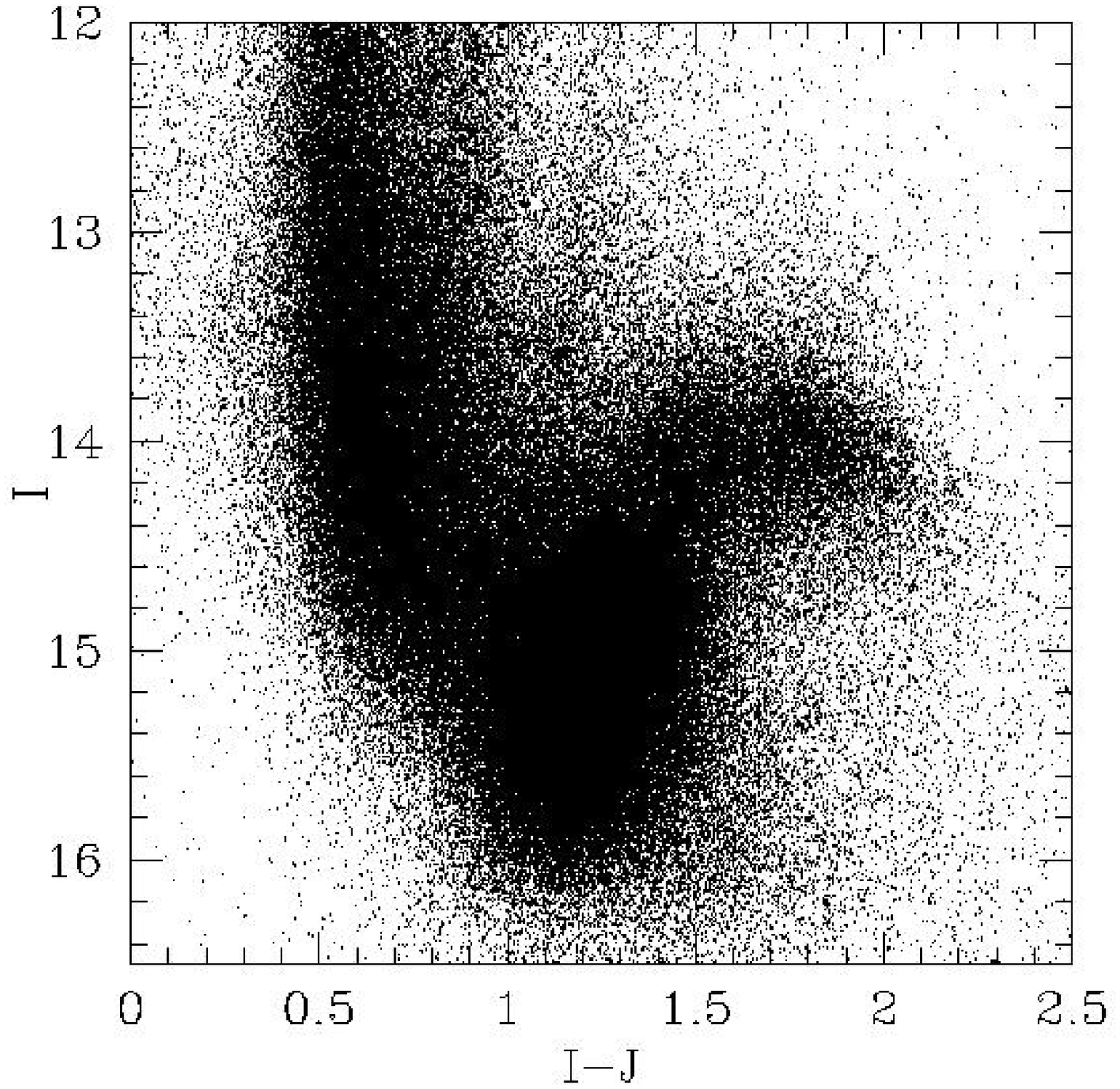}}
\end{minipage}\   \ \begin{minipage}[b]{6cm}
\resizebox{6cm}{!}{\includegraphics{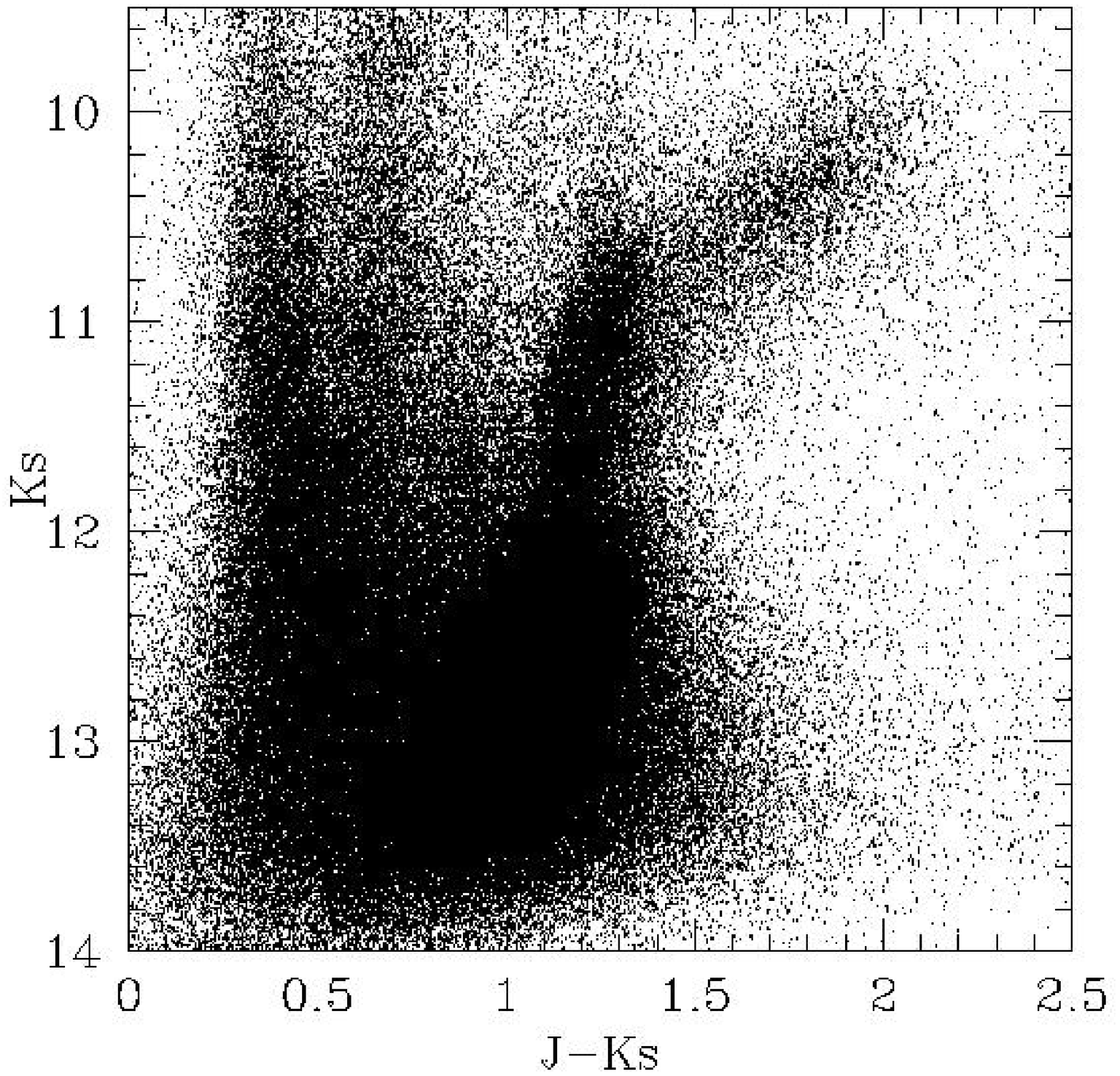}}
\end{minipage}
\caption{Colour--magnitude diagrams of DCMC sources detected simultaneously in three DENIS wave bands. Left: ($I-J$,$I$). The TRGB is at $I=14.54$. Right: ($J-K_S$, $K_S$), the TRGB is at $K_S=11.98$, O--rich AGB stars have $(J-K_S)=1.2$ and C--rich AGB stars have $(J-K_S)>1.4$.}
\label{cmds}
\end{figure}
In the following sections the  discussion is concentrated on the DENIS
measurements described  in my Ph.D. thesis (\cite{cionit}),  I am also
using EROS  and MACHO light--curves, and ISO  (LW2, LW10) measurements
covering small fields in the Clouds (Loup et al.~{\it in preparation}).

\section{The spatial distribution of AGB stars}
The  colour--magnitude diagrams (CMDs  -- Fig.~\ref{cmds})  made using
 the DENIS bands allow us to distinguish different groups of stars. In
 the  ($I-J$, $I$)  diagram the  densest region  is the  populous red
 giant branch of  the Large Magellanic Cloud (LMC).  It contains stars
 less massive then about $2$ solar masses and older than about $1.5-2$
 Gyr.  These  stars reach their maximum  luminosity at the  tip of the
 red  giant  branch (TRGB),  a  well defined 
 feature that  depends theoretically only  on the mass of  the stellar
 core.  The  plume of  objects brighter than  the TRGB are  AGB stars,
 which are  statistically separated into carbon (C--)  and oxygen rich
 (O--rich) in  the diagram ($J-K_S$, $K_S$) because  of the absorption
 effect  of the  different molecular  bands in  their  atmosphere (see
 figure  caption for  the  distinction).  Cepheids  and other  younger
 stars in the Clouds overlap the vertical sequences of our own Galaxy.
 These CMDs  made of sources detected  in the direction of  the LMC by
 DENIS  simultaneously in  all three  wave bands  show how  easily the
 different  stellar  groups  can  be  characterised  from  broad  band
 measurements.

\begin{figure}
\begin{minipage}[t]{6cm}
\resizebox{7cm}{!}{\includegraphics{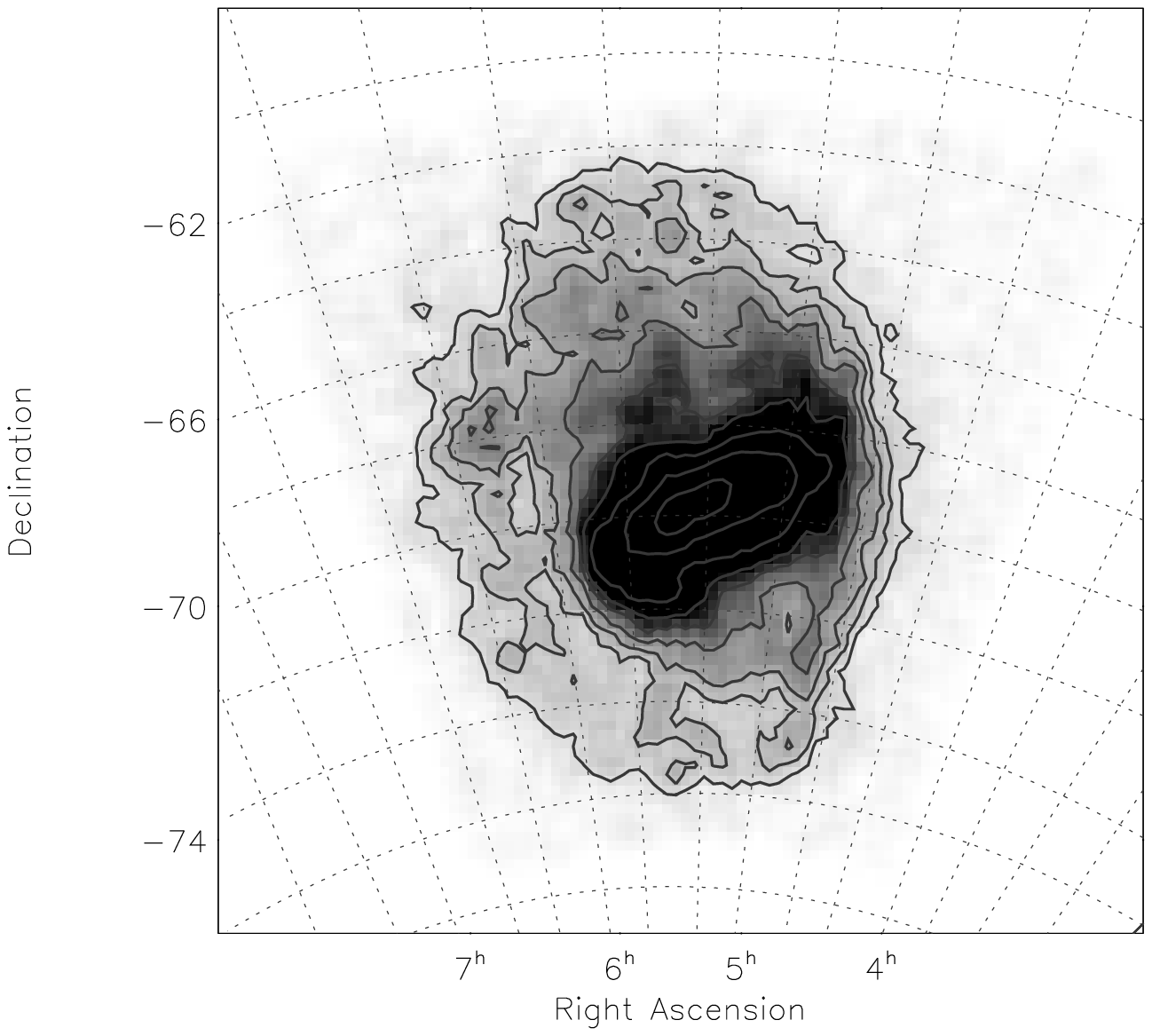}}
\end{minipage}\   \ \begin{minipage}[b]{6cm}
\resizebox{7cm}{!}{\includegraphics{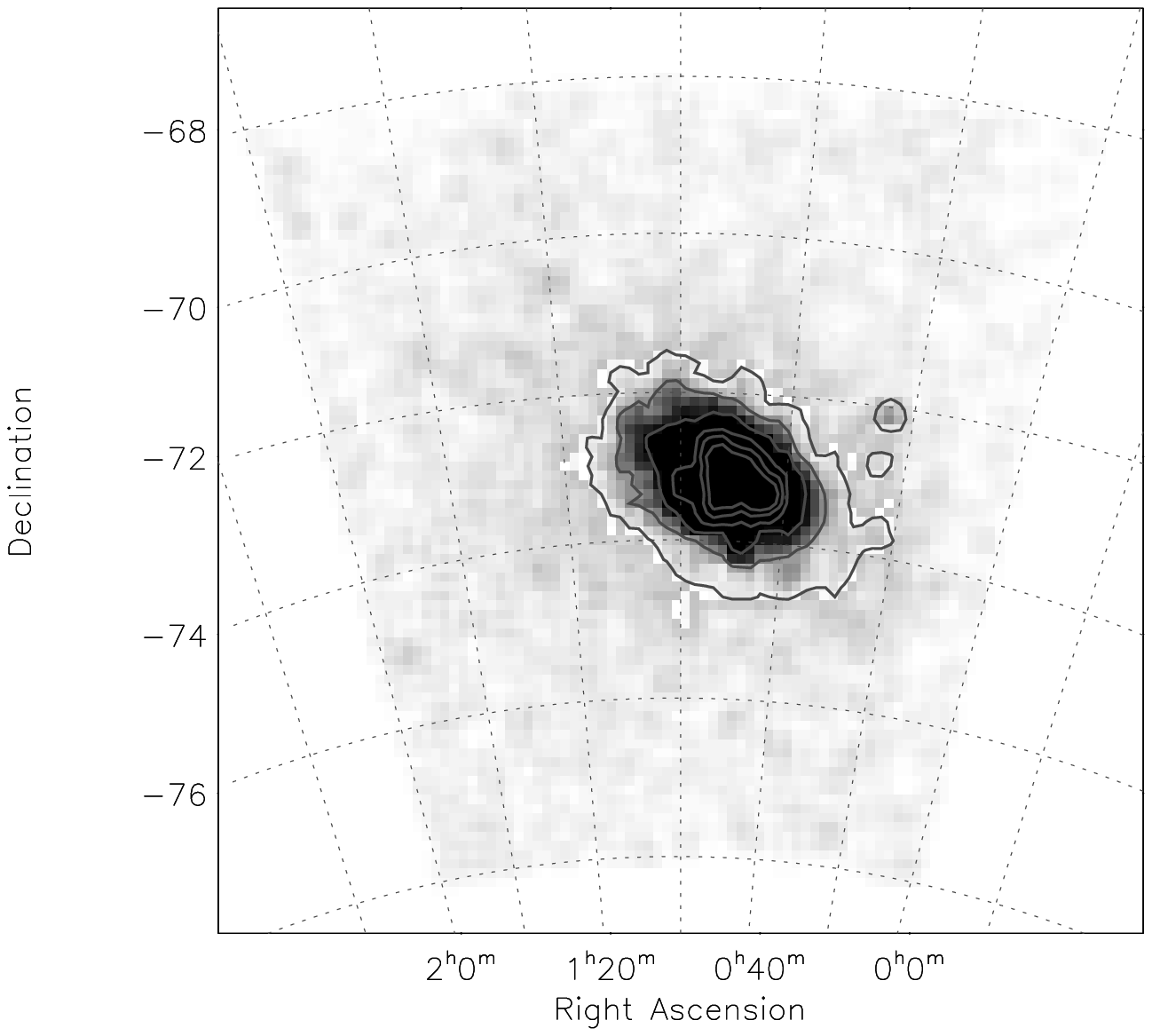}}
\end{minipage}
\caption{Star counts of LMC (left) and SMC (right) AGB stars in bins of 0.04 square degrees.}
\label{surface}
\end{figure}

If we select  AGB stars from the ($I-J$, $I$)  CMD, where the galactic
foreground contamination  is negligible, and  plot them in the  sky we
obtain an  extended smooth structure (Fig.~\ref{surface}).  
Note that the DCMC  is complete
for  optically thin AGB  stars; only  the most  obscured AGBs  are not
detected within the  DENIS limits.  In the LMC  we clearly distinguish
the bar and  an almost circular disk structure  of about $10^{\circ}$ 
in diameter. A clearly elliptical structure  is seen in the SMC with a
semi--major axis of about $3^{\circ}$  and a semi--minor axis of about
$2^{\circ}$.  Bins  are of  0.04  square  degrees.  See Cioni  et  al.
(\cite{cioni3}) for more details.

\begin{figure}
\begin{minipage}[t]{4cm}
\resizebox{\hsize}{!}{\includegraphics{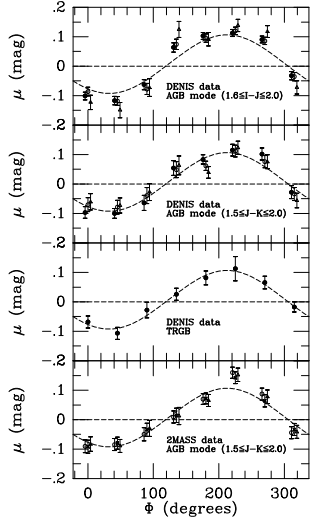}}
\end{minipage}\   \ \begin{minipage}[b]{3.5cm}
\caption{Variation of magnitude as a function of position angle in 8 disjunctive sectors between 2.5 and 6.7 degrees from the center of the LMC.}
\end{minipage}
\label{stru}
\end{figure}

If  we  consider the  shape  of  the LMC  and  imagine  to divide  the
 outermost part (ring that excludes the bar) in four sectors, then the
 modal  magnitude of a  selected sample  of stars  which span  a small
 interval  in magnitude  will  outline the  projected morphology.   In
 figure \ref{stru} as a function  of position angle it is shown the
 variation of the magnitude of,  from the top: AGB stars selected from
 the ($I-J$,  $I$) diagram  having $1.6<(I-J)<2.0$, C--rich  AGB stars
 selected from the  ($J-K_S$, $K_S$) diagram having $1.5<(J-K_S)<2.0$,
 the TRGB  obtained from the  magnitude distribution in the  $I$ band,
 and the C--rich AGB stars  selected form the ($J-K_S$, $K_S$) diagram
 within  the same colour  range as  before but  using this  time 2MASS
 data. There is one point per sector and per band. Results are in very
 good agreement.  The  dashed curve represents the best  fit model for
 an  inclination  angle of  $34.7^{\circ}$  and  a  position angle  of
 $122.5^{\circ}$.   The  sinusoidal  behaviour  shows  the  effect  of
 differences  in   distance  coupled  with  the   inclination  of  the
 Cloud. The  conclusion of  this work, described  in detail in  Van der
 Marel \& Cioni  (\cite{vdm1}) is that the intrinsic  shape of the LMC
 disk is elliptical  and it shows that AGB  stars are useful indicator
 of the extended structure of a galaxy.

\begin{figure}
\begin{minipage}[t]{6cm}
\resizebox{7cm}{!}{\includegraphics{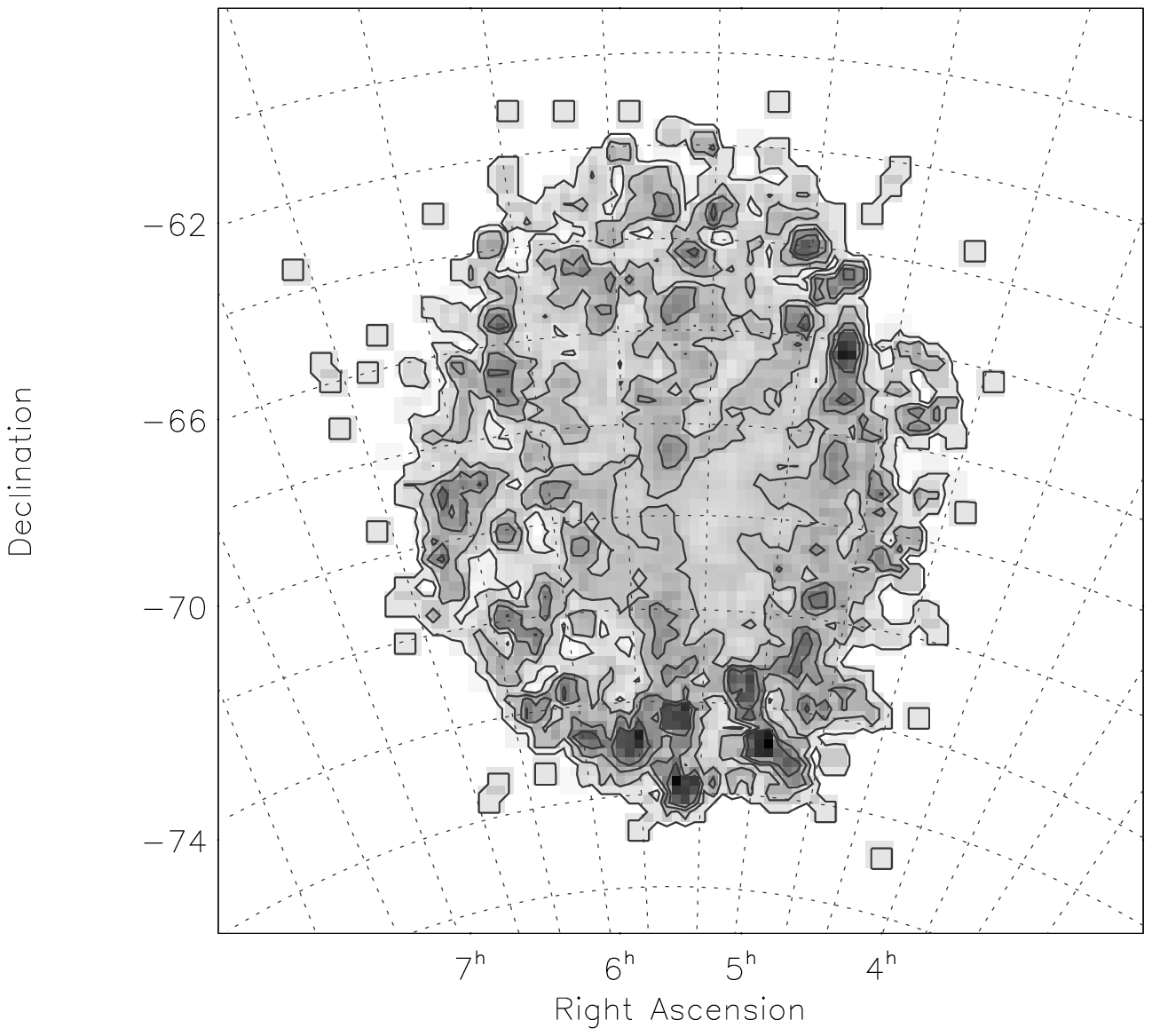}}
\end{minipage}\   \ \begin{minipage}[b]{6cm}
\resizebox{7cm}{!}{\includegraphics{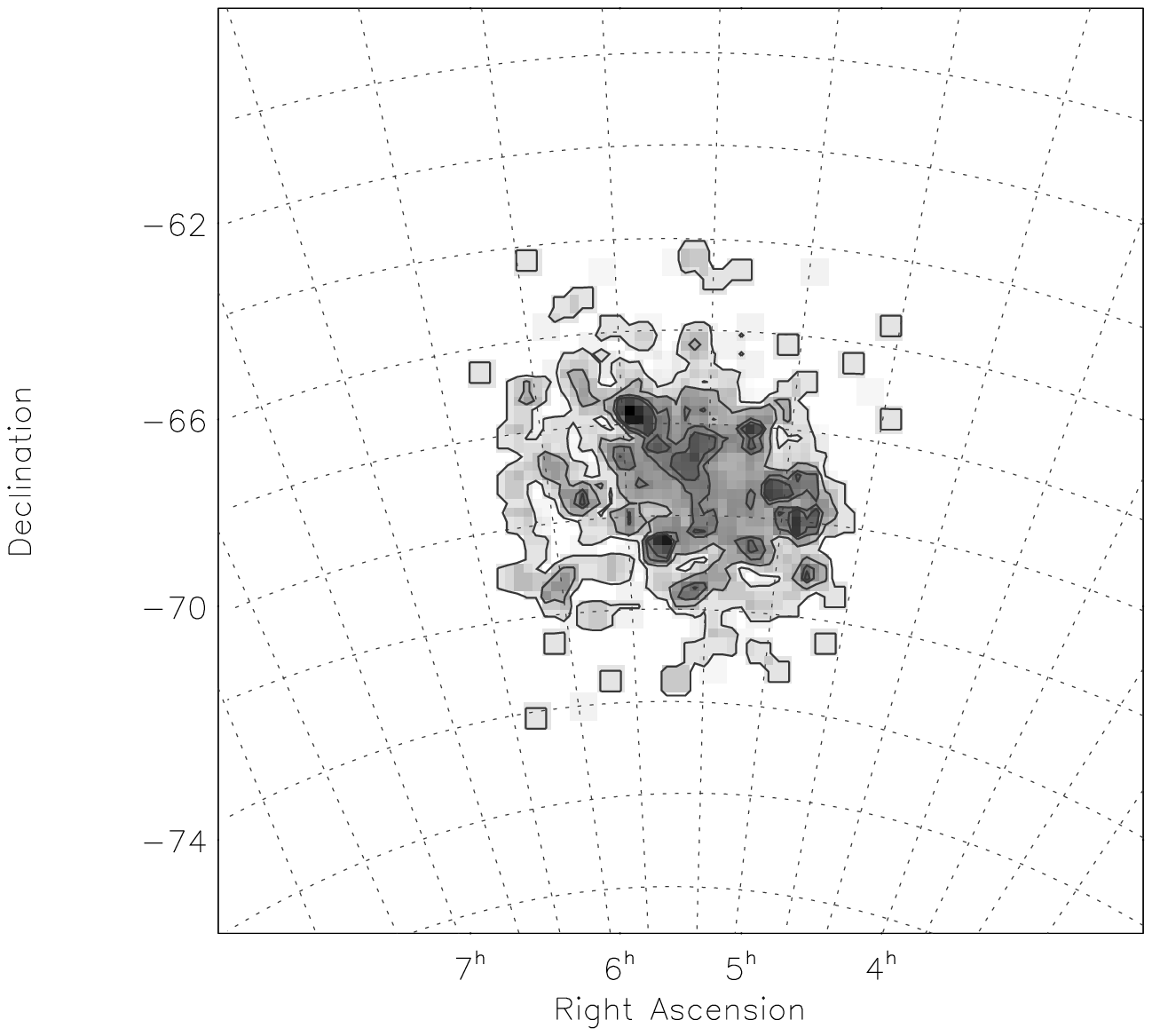}}
\end{minipage}
\caption{Distribution of the ratio between C--rich and O--rich AGB stars in the LMC (left) and in the SMC (right). Highest contours correspond to a C/M of $0.55$, the lowest to $0.1$}
\label{cmratio}
\end{figure}

The fact that there are AGB  stars both C-- and O--rich (M--type) 
is believed to
be primarily a  consequence of the third dredge--up  that enriches the
stellar atmosphere  with carbon products.  This dredge up seems  to be
more efficient in  metal--poor environments and that is why  we find more
C--rich AGB stars  in the SMC than  in the LMC and in  the Galaxy. The
C/M ratio can thus be used as a metalicity indicator. Figure \ref{cmratio}
shows the distribution
  of the C/M ratio  in both Clouds. The  separation between C--
and  O--rich  AGB  stars  has been  done  using  the  ($J-K_S$)
colour. The  darkest regions  correspond to a  higher C/M  ratio. 
In the  LMC there  are more  C--rich AGB  stars in  the external
parts,  thus  indicating  a  decreasing  metalicity  with  increasing
radius. In the SMC the situation  is more complicated and irregular 
(Cioni \& Habing, {\it in preparation}).
 
Using the VLT  telescope when the FLAMES instrument  will be available, 
the metalicity of  red giants could be measured more  precisely in
order  to study  the  chemical history  enrichment  of the  Magellanic
Clouds and the presence of a radial gradient. FLAMES is a multi--fibre
spectrograph which  will allow to observe $130$ optical spectra 
with  a resolution  of $\Delta \lambda / \lambda = 28000/17000$ 
and  up to $45000$  for the
eight fibres connected to the existing UVES instrument.

\section{The luminosity function ($K_S$)}
\begin{figure}
\resizebox{8cm}{!}{\includegraphics{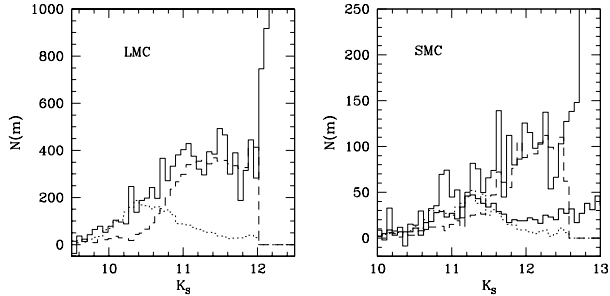}}
\caption{Histograms of the AGB stars in the LMC (left) and in the SMC (right). See text for details.}
\label{lumf}
\end{figure}
Fig.~\ref{lumf} shows the histogram of the magnitude of AGB stars. The
continuous  line indicates  the  whole distribution,  the dashed  line
indicates  only the O--rich  AGB stars  and the  dotted line  only the
C--rich AGB stars selected on the basis of their ($J-K_S$) colours. In
the  case of the  SMC there  is another  thicker continuous  line that
indicates  the  C--rich   AGB  stars  spectroscopically  confirmed  by
Rebeirot  (\cite{rebe}); this shows  that the  selection based  on the
colour is good enough for statistical studies. At the faintest mag the
sharp  increase of star  counts marks  the position  of the  TRGB. The
global shape of  the AGB histogram differs between  the Clouds  in
the sense  that it is  rather monotonically decreasing  for increasing
mag in the  SMC and it resembles  a bump, thus with a  flatter top, in
the LMC.   This difference in shape  can be attributed  to a different
star  formation history  between the  Clouds.  Habing derives the
following relation:
\begin{equation}
\frac{dL}{dM_{bol}}\propto SFR(t_{NOW}-t_{TO})\times \frac{dM_0}{dM_{bol}}
\end{equation}
 where the  slope of  the luminosity function  is proportional  to the
 star  formation rate  at  a  time $t_{TO}$  ago.  By considering  two
 different star  formation rates,  assuming the Salpeter  initial mass
 function and that  the stars decrease in $M_{bol}$  at the same rate,
 we  can  reproduce the  shape  of  the  luminosity functions  in  the
 Clouds.  We conclude that  for the  first $11$  Gyr of  our Universe
 there has been little formation of  stars in the LMC with mass higher
 than about $1.0$ solar mass. However, the SMC has older AGB 
 stars.  This is in  agreement with  the prediction  by Marigo  et al.
 (\cite{paola})  based on  the  study of  the  carbon star  luminosity
 functions.  Our  undergoing study  aims  at  deriving  the SFH  as  a
 function  of  position  in  the  Clouds (Habing  \&  Cioni,  {\it  in
 preparation}).

\section{AGB period--luminosity relation}
\begin{figure}
\begin{minipage}[t]{6cm}
\resizebox{\hsize}{!}{\includegraphics{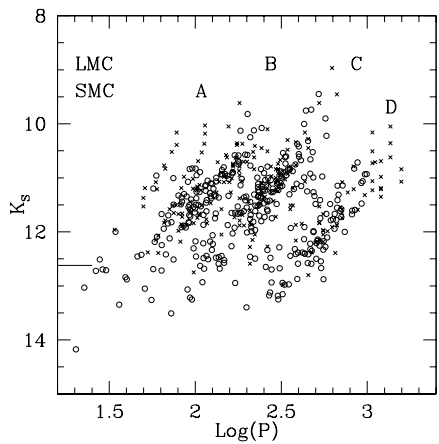}}
\end{minipage}\   \ \begin{minipage}[b]{4.5cm}
\caption{Period--magnitude diagram of long period variables in the Magellanic Clouds. Circles are for LMC stars and crosses for SMC stars.}
\end{minipage}
\label{var}
\end{figure}
Figure  \ref{var}  shows  the  period--magnitude  diagram  for
sources in the LMC and in the  SMC. The LMC sources are in the bar 
close to the optical center and  are   part   of   a  study   of
cross--identification   between  DCMC   sources  and   EROS2
light--curves (Cioni et al.~\cite{cioni4}).  
The sources  in the  SMC  were selected  from the  ISO
mini--survey catalogue of Loup et al.  (2002, soon to be released) to have a
confident   cross--identification   with    DCMC   or   2MASS.   Their
light--curves  were  obtained  from  the  MACHO  database.  The  first
conclusion we can deduce from  this figure  is that there  seem to  be no
clear differences between the two Clouds. Both sources occupy the four
slanted relations,  of which  relation C is  where Mira  variables are
also located. Note that all the 
sources have been placed  at the distance  of the SMC and that the 
$K_S$ magnitude is from DCMC.  The small
horizontal line indicates the  position of the TRGB. Sources occupying
relation D are multi-periodic and  the second period, also plotted, 
usually belong  to relation A  or B.  
In the SMC there are sources with longer periods, this is because 
 the MACHO light--curves cover a longer time range with respect to the 
EROS2 data. In addition there is a lack of SMC sources around the TRGB, 
this is probably  a selection effect because
these sources  do have an  emission at $7$  and $12$ micron.  ISO data
exist also for sources in the LMC and they are going to be used also to check 
for variables at the TRGB magnitude. 

\section{Summary} 
We  have seen that different groups  of red giants
can       be      distinguished       in       the      near--infrared
colour--magnitude--diagrams, we have also  seen that the AGB stars are
very good  indicators of the extended  structure of a  galaxy and that
from their luminosity function we  can obtain indications on the stars
formation history  and the metalicity,  the latter from the  ratio of
C-- over  O--rich AGB  stars. Finally, these  long period  variables 
follow the same period--luminosity relations in both Clouds.

\section{Asked questions}
{\bf P. Wood}. Do you take into account the variation of mass loss rate between the SMC and the LMC when you estimate the SFR from the AGB luminosity function?\\
{\bf Cioni} No, it is not taken into account. On the other hand these AGB 
stars are optically thin and do not loose mass at a very high rate, thus 
we do not expect a strong effect due to varying mass loss rate.

{\bf M. Jura}. For the LMC, is there any relationship in terms of ellipticity and/or positional angle of the ``elliptical'' disk and the inner bar?\\
{\bf Cioni} The inner bar requires a separate study which van der Marel is currently undertaking, thus I cannot at present answer your question which is indeed very interesting.

{\bf S. Deguchi}. How is the P--K relation of MSX sources?\\
{\bf Cioni} Among the sources discussed here there are very few MSX sources. The contribution by Wood discusses them in more detail.

\begin{acknowledgments}
I would like to kindly thank Cecile Loup for making available the ISO 
data prior publication and Harm Habing for a fruitful discussion and 
collaboration on the newest results. I am also grateful to M. Zoccali 
for comments on the text.
\end{acknowledgments}

\end{document}